\documentclass[]{article}
\usepackage{amssymb}
\usepackage{graphicx}
\usepackage{dcolumn}
\usepackage{bm}

\begin{document}
\title{Hydrostaticity and hidden order: effects of experimental conditions on the temperature-pressure phase diagram of URu$_{2}$Si$_{2}$}
\author{N. P. Butch\thanks{Present address: Center for Nanophysics and Advanced Materials, University of Maryland, College Park, MD 20742. Corresponding author. Email: nbutch@umd.edu}, J. R. Jeffries\thanks{Present address:  Lawrence Livermore National Laboratory, Livermore, CA 94550}, D. A. Zocco, and M. B. Maple
\\\vspace{6pt}
\small Dept. of Physics and IPAPS, University of California, San Diego, La Jolla, CA 92093
\\\vspace{6pt}
}
\maketitle
\begin{abstract}
The pressure-dependence of the hidden order phase transition of URu$_{2}$Si$_{2}$ is shown to depend sensitively upon the quality of hydrostatic pressure conditions during electrical resistivity measurements.  Hysteresis in pressure is demonstrated for two choices of pressure medium:  the commonly-used mixture of 1:1 Fluorinert FC70/FC77 and pure FC75. In contrast, no hysteresis is observed when the pressure medium is a 1:1 mixture of n-pentane/isoamyl alcohol, as it remains hydrostatic over the entire studied pressure range.  Possible ramifications for the interpretation of the temperature-pressure phase diagram of URu$_{2}$Si$_{2}$ are discussed.
\end{abstract}

\section{Introduction}
The heavy fermion superconductor URu$_{2}$Si$_{2}$ is notorious for undergoing a transition near $17$~K into a hidden order (HO) phase whose order parameter remains a matter of controversy some $20$ years since the compound's discovery. URu$_{2}$Si$_{2}$ crystallizes in the tetragonal ThCr$_{2}$Si$_{2}$ structure with lattice parameters of $4.13$~\AA\ and $9.58$~\AA; this structural anisotropy is reflected in the bulk transport and magnetic properties \cite{Palstra85,Schlabitz86}.  The transition from the paramagnetic (PM) to HO phase is marked by a large specific heat anomaly at $17$~K and appears to be associated with a Fermi surface instability \cite{Maple86,Behnia05,Sharma06}.  In the HO phase, a small staggered moment is observed, but its magnitude cannot account for the large entropy liberated by the HO transition \cite{Broholm87}, leading to speculation that the entropy is instead associated with some hidden order parameter, whose nature still remains elusive.  Similarly unresolved  is the relationship between the HO phase and the superconducting (SC) state found below $1.5$~K.

In recent years, pressure studies of the HO phase have become increasingly common as more techniques have been developed for reliable use under hydrostatic pressure, leading to an apparent convergence towards a common  temperature-pressure ($T-P$) phase diagram for URu$_{2}$Si$_{2}$. However, an important point that appears not to have received enough attention is the degree of hydrostaticity of the pressure environments in which these measurements have been performed, and how the evolution of the HO state is affected by these pressure conditions. The present study addresses this issue in the context of electrical resistivity measurements.

Despite two decades of research, current understanding of the $T-P$ phase diagram of URu$_{2}$Si$_{2}$ remains incomplete.  Early measurements established that the HO transition temperature $T_{0}$ increases under applied hydrostatic pressure \cite{deBoer86,Louis86,Onuki87,Fisher90} and that the rate of change $\frac{\partial T_{0}}{\partial P}$ increases substantially at a critical pressure $P_{c} \approx 12$~kbar \cite{Maple86,McElfresh87,Uwatoko92,Ido93,Brison94}. In contrast, the SC critical temperature $T_{c}$ was found to be suppressed to $0$~K at a similar $P_{c}$. Via neutron diffraction, the small staggered moment was observed to grow by an order of magnitude under pressure, giving way to large-moment antiferromagnetic (AFM) order at $15$~kbar \cite{Amitsuka99}.  Evidence for the heterogeneous coexistence of the HO and AFM states below $15$~kbar was suggested from $^{29}$Si NMR measurements \cite{Matsuda01,Matsuda03} and $\mu$SR measurements indicated that the AFM volume fraction emerged at $\approx 5$~kbar (extrapolated to $T=0$~K) \cite{Amitsuka03,Amato04}.  This general pressure-dependence was corroborated by measurements of dilation under pressure \cite{Motoyama03}, while measurements of ac magnetic susceptibility demonstrated the suppression of the SC state at approximately $4$~kbar, at the $T=0$~K onset of the HO-AFM transition \cite{Uemura05,Sato06}.

Two important current lines of inquiry regarding the URu$_{2}$Si$_{2}$ $T-P$ phase diagram are how the HO-AFM phase boundary relates to $1)$ the higher-temperature PM-HO/AFM boundary, which might indicate whether HO and AFM coexist microscopically \cite{Bourdarot04,Mineev05}, and $2)$ the lower-temperature SC phase boundary. While some neutron scattering reports concluded that the HO-AFM boundary is separate from the PM-HO/AFM curve \cite{Bourdarot04,Bourdarot05}, a more recent report of neutron scattering and ac susceptibility suggested a sharp transition at $7$~kbar, consistent with HO-AFM phase separation \cite{Amitsuka07,Matsuda07}, and supported a $\sim 4$~kbar SC critical pressure. The latter phase diagram was recently corroborated by measurements of electrical resistivity and ac calorimetry \cite{Hassinger08}.  However, several details regarding the HO-AFM phase boundary are unresolved: the de Haas-van Alphen effect appears insensitive to the HO-AFM transition \cite{Nakashima03}, as does the high-magnetic-field phase diagram \cite{Jo07}, while the SC phase has also been observed to persist at pressures significantly greater than  $5$~kbar in magnetization \cite{Tenya05} and electrical resistivity measurements \cite{Jeffries07,Jeffries08}.

A survey of publications indicates that the following media have been used in studies of URu$_{2}$Si$_{2}$: helium \cite{deBoer86,Louis86,Bourdarot05}, 1:1 n-pentane/isoamyl alcohol \cite{McElfresh87,Brison94,Amato04,Jeffries07,Jeffries08}, transformer oil/kerosene \cite{Uwatoko92,Kagayama94,Oomi94,Kagayama98}, 1:1 FC70/FC77 \cite{Amitsuka99,Matsuda01,Matsuda03,Motoyama03,Uemura05,Sato06,Bourdarot05,Amitsuka07,Matsuda07,Kagayama94,Oomi94,Kagayama98,Nishioka00,Amitsuka08}, 4:1 methanol/ethanol \cite{Pfleiderer06}, argon \cite{Hassinger08}, and Daphne oil \cite{Amitsuka08}.  In some reports, the pressure medium was not disclosed \cite{Onuki87,Fisher90,Ido93,Amitsuka03,Bourdarot04,Nakashima03,Tenya05,Knebel07,Jo07}. It is remarkable that through the recent development of the $T-P$ phase diagram of URu$_{2}$Si$_{2}$, one fact has remained mostly overlooked: the pressure studies upon which the phase diagram are based are almost exclusively performed using a 1:1 mixture of Fluorinert FC70 and FC77 that is hydrostatic only below a modest pressure. In fact, the range of pressure in which the HO-AFM transition has been identified corresponds to the room temperature limit of hydrostaticity of the mixture, beyond which the medium starts to support shear stresses, a condition that often leads to unexpected effects in a material.

The Fluorinert family of fluorinated hydrocarbons (3M St. Paul, MN) are liquids commonly used in pressure work, due to their ease of use and low reactivity. An extensive study of the hydrostaticity of these compounds has been carried out under pressure, demonstrating that the room-temperature hydrostatic limit of a 1:1 mixture of FC70/77 is $8$~kbar \cite{Sidorov05,Varga03}.  Among the different fluid media available for hydrostatic pressure experiments, helium is regarded as closest to ideal, remaining hydrostatic up to at least  $500$~kbar at room temperature \cite{Kenichi01}. Its usage, however, is limited because of the experimental difficulty typically associated with loading and containing it in a pressure cell.  The hydrostatic limits of other commonly-used media have also been determined: the mixture of 1:1 n-pentane/isoamyl alcohol was shown to remain hydrostatic at room temperature to pressures greater than $30$~kbar \cite{Jayaraman67}, Daphne oil 7373 to $20$~kbar \cite{Yokogawa07,Fukazawa07}, 4:1 methanol/ethanol to $98$~kbar \cite{Angel07}, and argon to $19$~kbar \cite{Angel07}. Note that these are room temperature values, and that at pressures lower than those listed, the media can solidify at temperatures less than $300$~K.  If the applied pressure is changed when the medium is a fluid, shear stresses will be small if the fluid solidifies when it is cooled. However, if pressure is applied to a solid medium, significant shear stresses will be induced.  The magnitude of this effect is sensitive to experimental details including the geometry of a pressure cell.  For example, anomalies in data at pressures appreciably less than $20$~kbar have been ascribed to the solidification of Daphne oil 7373 \cite{Isotalo95}.

The effects of the hydrostatic pressure medium on measurements of URu$_{2}$Si$_{2}$ have only been reported in two publications. In the first, it was noted that neutron scattering measurements performed in He, which were limited to less than $5$~kbar, showed no sign of AFM moment, while those performed in Fluorinert in order to access higher pressures already detected a moment at $4.5$~kbar \cite{Bourdarot05}. Recently, a comparison of neutron scattering and ac susceptibility measurements in  Fluorinert FC70/FC77 and Daphne oil 7373 was reported, which found that the HO-AFM transition and disappearance of SC both occurred at slightly higher pressure in the Fluorinert \cite{Amitsuka08}. Arguing that a loss of hydrostaticity in Fluorinert would increase stress along the c-axis and increase $T_{c}$, the authors asserted that Daphne oil provided a better hydrostatic pressure environment. These rudimentary investigations clearly demonstrate the need to properly characterize the effects of non-hydrostaticity on the physical properties of URu$_{2}$Si$_{2}$.

Electrical resistivity measurements continue to play an integral role in the construction of $T-P$ phase diagrams of URu$_{2}$Si$_{2}$ despite the fact that the recent evolution in our understanding  has been motivated largely by neutron scattering, NMR, and $\mu$SR studies. This is primarily because the observation of an ordered moment in neutron scattering, NMR, and $\mu$SR experiments often occurs at temperatures that do not match $T_{0} \approx 17$~K estimated from resistivity and specific heat data.  In order to better understand the ramifications of the loss of hydrostaticity in Fluorinert, the present investigation was carried out to compare electrical resistivity studies of the PM-HO/AFM transition performed in two Fluorinert mixtures and in 1:1 n-pentane/isoamyl alcohol, which remains hydrostatic over the entire range of interest.  It has been found that the measurements performed in Fluorinert exhibit significant hysteresis in pressure, whereas the data collected using 1:1 n-pentane/isoamyl alcohol exhibit no hysteresis.  The consequences of these results, specifically in regard to interpretation of the established temperature-pressure phase diagram of URu$_{2}$Si$_{2}$, are discussed.

\begin{figure}[tbp]
    \begin{center}
    \rotatebox{0}{\includegraphics[width=3.375in]{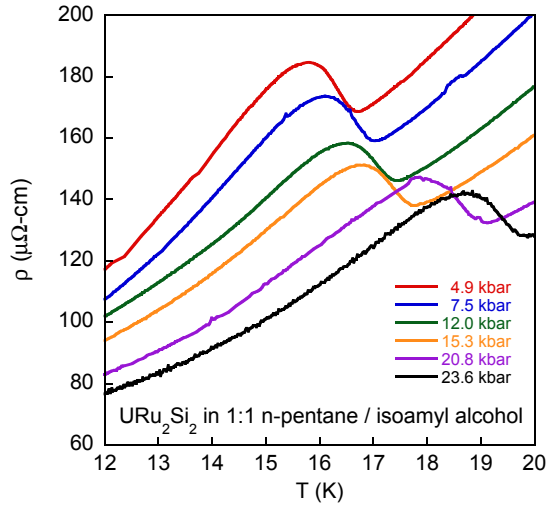}}
    \end{center}
    \caption{Electrical resistivity versus temperature in the vicinity of the hidden order transition at various pressures using 1:1 n-pentane/isoamyl alcohol as the pressure transmitting medium.}
    \label{rhoT}
\end{figure}

 \section{Experiment}
Single crystals of URu$_{2}$Si$_{2}$ were synthesized via the Czochralski technique in a tri-arc furnace under flowing argon. Samples were annealed at $900$~$^{\circ}$C for $7$ days in the presence of a Zr getter. Phase homogeneity was verified by powder x-ray diffraction on ground pieces of the single crystals.  Samples for electrical resistivity measurements were oriented via the back-reflection Laue method and spark cut and gently sanded to size.  The pressure dependence of the electrical resistivity along the $[100]$ direction was measured in BeCu piston-cylinder clamps to pressures of approximately $25$~kbar. The samples were placed in a teflon capsule filled with one of three pressure-transmitting liquids: Fluorinert FC75 (3M, St. Paul, MN), a 1:1 mixture by volume of Fluorinert FC70 and FC77 (3M, St. Paul, MN), and a 1:1 mixture by volume of n-pentane (Fisher, Fair Lawn, NJ) and isoamyl alcohol (EMD, Gibbstown, NJ), hereafter referred to as NP/IA for brevity. Two URu$_{2}$Si$_{2}$ samples were used in this study, one sample in FC75, which was destroyed by a failed pressure measurement, and one sample in  the NP/IA and FC70/77.  The applied pressure was adjusted at room temperature.  Temperature $T$ was determined via a calibrated Cernox thermometer attached to the cell, while inductive measurements of the superconducting transition of Pb were used as a manometer. Uncertainties in pressure for all measurements were less than $\pm0.5$~kbar.  Electrical resistivity $\rho$ of the samples was measured using an LR-700 ac resistance bridge with currents of less than $100 \mu$A. All measurements were carried out in a $^{4}$He cryostat.  The electrical properties of the two samples were comparable: the FC75 sample had a residual resistivity ratio (RRR) of $12$ and $T_{0}(1$~bar$)=17.0$~K, while the other sample had a RRR of $7$ and $T_{0}(1$~bar$)=16.4$~K. Results of the experiment in NP/IA have been published \cite{Jeffries07,Jeffries08}.

\section{Results and discussion}
\begin{figure}[tbp]
    \begin{center}
    \rotatebox{0}{\includegraphics[width=6.0in]{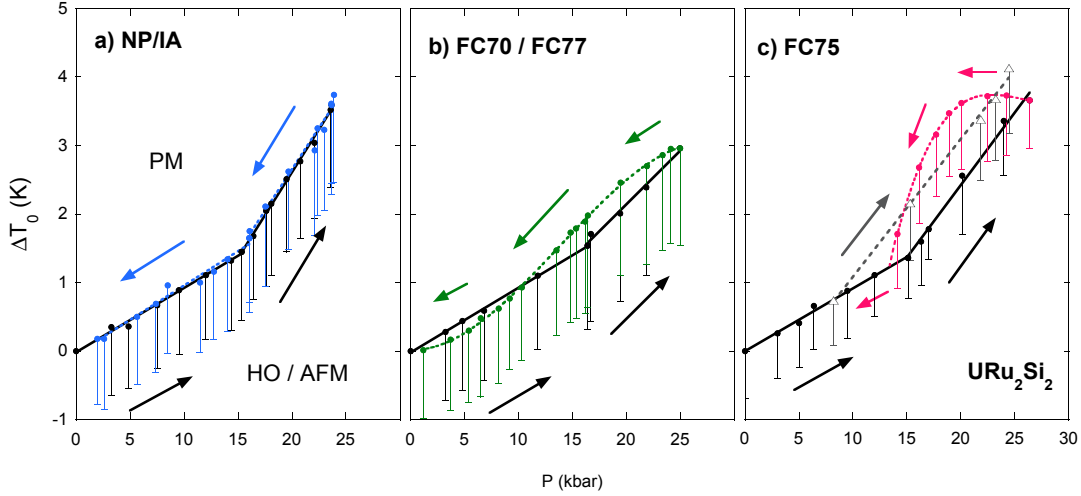}}
    \end{center}
    \caption{Temperature-pressure phase boundary between paramagnetic (PM) and hidden order (HO) / antiferromagnetic (AFM) phases, based on electrical resistivity measurements in different pressure transmitting media. a) 1:1 mixture of n-pentane/isoamyl alcohol.  No hysteresis in pressure is observed. b) 1:1 mixture of Fluorinert FC70/FC77.  The high pressure slope is less than that seen in NP/IA and hysteresis is evident upon depressurization. c) Fluorinert FC75. While the curve upon pressurization is similar to the one in NP/IA, significant hysteresis is observed upon depressurization and re-pressurization (open triangles).}
    \label{PT0}
\end{figure}

The resistive anomaly associated with the URu$_{2}$Si$_{2}$ hidden order transition takes the form of a small narrow peak in $\rho(T)$ at $T_{0}$(1~bar)$\approx 17$~K. Examples of this transition, measured at different pressures $P$ in NP/IA, are illustrated in Figure~\ref{rhoT}. This general structure is maintained over the entire pressure range studied. The three $T-P$ phase boundaries separating the PM, HO, and AFM phases, determined using different hydrostatic media are plotted in Figure~\ref{PT0}. The points are defined by the local minima in $\rho(T)$, while the vertical bars indicate the width in $T$ of the transition, with the lower endpoint corresponding to the local maximum in $\rho(T)$. It is evident that in all three pressure media, the transition width, roughly $1$~K from resistive maximum to minimum, does not change appreciably. The arrows indicate whether pressure is being applied or released. For ease of comparison between phase boundaries, the change in the HO transition temperature relative to its zero pressure value $\Delta T_{0}(P)=T_{0}(P)-T_{0}(1$~bar$)$ is presented.

The phase boundary constructed from measurements in NP/IA (Figure~\ref{PT0}a) is presented first, as it is reversible and reflects the most hydrostatic environment.  Its most striking feature is a discontinuous change in $\frac{\partial T_{0}}{\partial P}$ at $P_{c} \approx 15$~kbar, associated with the transition from HO to long range AFM.  For $P<15$~kbar,  $\frac{\partial T_{0}}{\partial P} = 0.094$~K~kbar$^{-1}$, while for $P>15$~kbar,  $\frac{\partial T_{0}}{\partial P} = 0.25$~K~kbar$^{-1}$.  Most importantly, there is no difference in $\frac{\partial T_{0}}{\partial P}$ whether increasing or decreasing $P$. The room-temperature hydrostatic limit of NP/IA has been shown to be at least $30$~kbar \cite{Jayaraman67}, which exceeds the pressure range of the present study. The sharp change in slope at $15$~kbar and lack of hysteresis attest to the hydrostaticity of this mixture, in contrast to the phase boundaries considered next.

The phase boundary constructed from measurements in FC70/FC77 (Figure~\ref{PT0}b) is superficially similar to that in Figure~\ref{PT0}a. However, there is a difference in slope at high pressure and the kink at $15$~kbar upon pressurization is much less pronounced.  For $P<15$~kbar,  $\frac{\partial T_{0}}{\partial P} = 0.094$~K~kbar$^{-1}$, while for $P>15$~kbar,  $\frac{\partial T_{0}}{\partial P} = 0.16$~K~kbar$^{-1}$, which is significantly lower than in Figure~\ref{PT0}a.  The depressurization curve lacks an abrupt kink in slope and instead has a gentle \emph{S} shape, crossing the pressurization curve at roughly $11$~kbar.  Following the analysis of the data taken upon pressurization, for $P<15$~kbar, a fit with two straight lines to the points taken upon depressurization yields an intersection in the vicinity of $7-8$~kbar.  The weakened $\frac{\partial T_{0}}{\partial P}$ at higher $P$, and the smooth depressurizing $T-P$ curve can be explained by the loss of hydrostaticity at about $8$~kbar \cite{Sidorov05}. To check whether the observed hysteresis could be due to the creation of permanent strains in the URu$_{2}$Si$_{2}$ sample through cold working via repeated quasi-hydrostatic compression, the same sample was measured a second time in NP/IA after the measurement in FC70/77. As the $T-P$ phase boundary in NP/IA was unchanged, it appears that the observed hysteresis can be entirely ascribed to the pressure medium.

Measurements in FC75 yield a phase boundary (Figure~\ref{PT0}c) dramatically different from the previous two. Initially, the pressurization curves closely match those from the measurement in NP/IA and a sharp kink is visible at $15$~kbar.   For $P<15$~kbar,  $\frac{\partial T_{0}}{\partial P} = 0.092$~K~kbar$^{-1}$, while for $P>15$~kbar,  $\frac{\partial T_{0}}{\partial P} = 0.21$~K~kbar$^{-1}$.  The most noticeable feature of this phase boundary is the dramatic  hysteresis upon initial depressurization.  This curve eventually meets the pressurization curve at about $12$~kbar, which corresponds to the hydrostatic limit of FC75 \cite{Sidorov05}.  The next point, at $8$~kbar, lies on the low-$P$ pressurization curve, but once pressure is increased again, the data appear to fall on another line (open triangles).  The slope of this line, $\frac{\partial T_{0}}{\partial P} = 0.20$~K~kbar$^{-1}$, matches the initial high-$P$ pressurization slope, although the data are shifted to lower pressure by about $3$~kbar.  Thus, even though the $8$~kbar point appears to fall on the initial pressurization curve, the original high-$P$ behavior is not recovered, indicating that residual shear stresses persist.  It is also noteworthy that, despite its nominally higher hydrostatic limit relative to FC70/FC77, the hysteresis observed in FC75 is much more pronounced.  This unexpected result indicates that the limits of hydrostaticity are not necessarily reliable predictors of the magnitude of non-hydrostatic effects in a pressure experiment, and may reflect a difference in the softening of the two media upon depressurization.

\begin{table}
\caption{Summary of electrical resistivity data measured in 1:1 n-pentane/isoamyl alcohol (NP/IA), 1:1 Fluorinert FC70/FC77, and Fluorinert FC75. $^{\rm a}$From References \cite{Sidorov05} and \cite{Jayaraman67}}
\begin{center}
{\begin{tabular}{lccc}
 \hline
& NP/IA & FC70/FC77 & FC75\\
\hline
\textbf{Pressurization} & & & \\
$P_{c}$ (kbar) & $15$ & $15$ & $15$\\
$P<P_{c}$ $\frac{\partial T_{0}}{\partial P}$ (K~kbar$^{-1}$) & $0.094$ & $0.094$ & $0.092$\\
$P>P_{c}$ $\frac{\partial T_{0}}{\partial P}$ (K~kbar$^{-1}$) & $0.25$ & $0.16$ & $0.21$\\
\textbf{Depressurization} & & & \\
Hysteresis & none & \emph{S} shape & dramatic\\
$P_{c}$ (kbar) & $15$ & $7-8$ & $12$\\
\hline
Hydrostatic limit$^{\rm a}$ (kbar) & $>30$  & $8$  & $12$ \\
\hline
\end{tabular}}
\end{center}
\label{tbl:compare}
\end{table}

Metallurgical defects in URu$_{2}$Si$_{2}$ have been suggested to lead to the sample-dependence of some physical properties \cite{Ramirez91,Fak96,Amitsuka07}, but they do not significantly affect the PM-HO/AFM transition curve.  In the present study, the RRR of the sample measured in NP/IA and FC70/FC77 is modest in relation to that of a recently reported sample \cite{Kasahara07}, although it is closer to the values reported elsewhere.  Despite the implied magnitude of defects in the sample used in the present study, there is a sharp signature at $P_{c} \approx 15$~kbar in NP/IA, while the $\frac{\partial T_{0}}{\partial P}$ change is much less obvious in FC70/FC77 and is even shifted to lower $P$ upon depressurization. Furthermore, the repetition of the NP/IA measurement after the FC70/FC77 measurement demonstrates that any strains induced in the sample are not permanent and that hysteresis is attributable to the pressure medium.  It is evident that the sharpness of the phase boundaries, even for $P_{c} < 10$~kbar, is determined almost entirely by the pressure medium employed, not sample-dependence.

The extreme sensitivity of URu$_{2}$Si$_{2}$ to  inhomogeneities in the pressure medium is consistent with the results of uniaxial stress measurements. From early studies, it is clear that URu$_{2}$Si$_{2}$ exhibits a strong dependence on the direction of applied stress: stress applied along the $a$ axis suppresses $T_{c}$ and enhances $T_{0}$, while stress applied along the $c$ axis has the opposite effect of enhancing $T_{c}$ and suppressing $T_{0}$ \cite{Bakker92,Guillaume99}. Neutron scattering experiments under uniaxial stress have demonstrated that the small ambient-pressure AFM moment increases only when stress is applied along the basal plane, not along the $c$ axis \cite{Yokoyama05}.

In the present study, although the hydrostatic limits were exceeded upon pressurization, the initial increasing-pressure curves of all three media are qualitatively similar, with kinks at $P_{c} \approx 15$~kbar, while the biggest differences occur upon removal of pressure.  Because it is not uncommon to record data only upon depressurization of a cell, the opportunity for error is appreciable.  The $T-P$ data taken in FC75 upon depressurization look obviously suspicious and would likely be recognized as erroneous. In contrast, if one only looked at the depressurization data of FC70/FC77 without having taken data upon pressurization, it would appear that $P_{c} \approx 7$~kbar instead of $15$~kbar.  Given that the low-$T$ portion of the HO-AFM boundary is observed at about $7$~kbar in neutron scattering, NMR, and $\mu$SR studies, a $7$~kbar $P_{c}$ at higher $T$ could appear to be reasonable. An incorrect, low value of $P_{c}$ deduced from transport measurements could be easily misinterpreted, leading to the conclusion that the HO-AFM phase boundary is almost vertical. If non-hydrostatic effects similarly influence other types of measurements, it could also potentially force the entire HO-AFM boundary to appear to occur at lower $P$ than it really does.

While neutron scattering, NMR, and $\mu$SR probes are sensitive to the HO-AFM transition, in order to generate a full phase diagram, it has been necessary to also include the PM-HO/AFM transition determined from electrical resistivity data \cite{Amato04,Bourdarot04,Bourdarot05,Amitsuka07,Matsuda07,Amitsuka08,Knebel07}.  Conclusions made about the coupling between HO and AFM parameters depend upon extrapolations between these two curves, yet the resistivity data are rarely discussed beyond inclusion in the phase diagram. In these reports, a cursory inspection of the phase boundaries based upon electrical transport studies exposes a similarity to the depressurization curve of Fluorinert FC70/FC77 in Figure~\ref{PT0}b, lacking a sharp discontinuity in $\frac{\partial T_{0}}{\partial P}$, and yielding a relatively low value of $P_{c}$ where one can be estimated.  Exhibiting similar features, these phase diagrams are consistent with those based upon thermal expansion and ac magnetic susceptibility measurements, also performed in FC70/FC77 \cite{Motoyama03,Uemura05,Matsuda07}.  It was only recently that measurements of electrical resistivity and ac calorimetry performed in Ar have been able to corroborate this general phase diagram \cite{Hassinger08}. However, it is important to note that despite the higher limit of hydrostaticity of Ar relative to FC70/FC77, the pressure in \cite{Hassinger08} was adjusted at low temperature; because Ar freezes below $85$~K at $1$~bar, the medium could not have been fluid.

The $T-P$ phase diagrams from the present study are consistent with the published limits of hydrostaticity of the respective pressure media and underscore the need to study the properties of URu$_{2}$Si$_{2}$ under carefully monitored pressure conditions. The evidence presented herein suggests that the discrepancies between reported $T-P$ phase diagrams of URu$_{2}$Si$_{2}$ may be predominantly due to the way in which a sample was measured, not because the properties of the sample are intrinsically different.  The hysteretic behavior observed in this study would likely be most similar  in measurements sensitive to the same phenomena as electrical transport in URu$_{2}$Si$_{2}$, i.e., bulk property measurements such as magnetization and specific heat.  Short of revisiting all published investigations performed using non-hydrostatic pressure media, in which the data may have been collected upon only pressurization or depressurization, the next best approach is to suggest the performance of these studies under controlled hydrostatic conditions. For example, because no moment was observed up to $5$~kbar in neutron scattering experiment utilizing He \cite{Bourdarot05}, NMR and $\mu$SR measurements performed in He should also determine no AFM volume fraction below at least $5$~kbar.

\section{Summary}

The present study has demonstrated a significant dependence of the quality of electrical resistivity measurements of the hidden order transition of URu$_{2}$Si$_{2}$ on the pressure medium used.  A 1:1 mixture of n-pentane/isoamyl alcohol provided hydrostatic pressure conditions in the entire pressure range studied, up to roughly $26$~kbar, while the Fluorinert liquids FC75 and a 1:1 mixture of FC70/FC77 exhibited hysteresis consistent with their solidification under pressure.  The effects of non-hydrostaticity may play a significant role in the differences between temperature-pressure phase diagrams in the literature, and they should be interpreted with this in mind.  As the pressure-dependent properties of URu$_{2}$Si$_{2}$  have been shown to be sensitive to non-hydrostatic conditions, careful consideration of the pressure medium and its hydrostatic limits are of paramount importance with regard to comparing results over a wide range of pressures.

\section*{Acknowledgments}

The authors thank S. K. McCall and D. D. Jackson for helpful discussions.  This research was supported by the National Nuclear Security Administration under the Stewardship Science Academic Alliances program through DOE Research Grant~\#~DE-FG52-06NA26205.

\end{document}